# SANAYİ ÜRETİMİNİN EKONOMİK BÜYÜME ÜZERİNDEKİ ETKİSİ:

# MADDİ KALKINMA ÇERÇEVESİNDE TÜRKİYE İÇİN YENİ AMPİRİK KANITLAR

Araş. Gör. Ali DOĞDU

Doğu Akdeniz Üniversitesi

ORCID ID:

https://orcid.org/0000-0003-0556-8255

Doç. Dr. Murad KAYACAN

Uluslararası Final Üniversitesi

ORCID ID:

https://orcid.org/0000-0002-7606-6183

**Özet**

İnsanlığın gelişimi ile devletlerin oluşumu sonrası birçok farklı süreç atlatmış olan ekonomi literatürü, temel noktada benimsemiş olduğu hedefler açısından en önemli dayanak olarak ekonomik büyüme ve kalkınmayı ele almıştır. Ekonomik büyüme ve kalkınma için literatürde farklı görüşler bulunmaktadır. Büyüme ve kalkınma hipotezleri birçok farklı açıdan ele alınmıştır. Tarihsel ve teorik olarak süreç içerisinde farklı yöntem ve araçlarla bu süreci yönetmeyi amaçlamışlardır. Gelişmiş ülkelerin sanayileşme süreçleri dikkate alındığında kalkınma ve büyümenin önemi gelişmekte olan ülkelere referans niteliğinde olmaktadır. Türkiye içinde ekonomik büyüme ve kalkınma Cumhuriyetin ilk yıllarından bu yana üzerinde durulan konuların başında gelmektedir. Türkiye'de Prof. Dr. Necmettin Erbakan tarafından Maddi Kalkınma kapsamında önerilen Ağır Sanayi hamlesi ile üretim yapılması görüşünün incelenmesi amaçlanmıştır. Türkiye ekonomi tarihinde gerçekleştirilen farklı kalkınma stratejileri içerisinde en önemlisini teşkil eden beşer yıllık dönemler için planlarına kalkınma planları zaman içerisinde yerini kısa-orta ve uzun vadeli kalkınma planlarına bırakmıştır. Kalkınmanın temel taşlarından birini ise Erbakan'ın da dediği iktisadi büyüme süreçleri oluşturmaktadır. Erbakan'a göre bu iktisadi büyüme kalkınmayı sağlayarak ülkenin gelişmiş ülke konumuna ulaşmasını sağlayacaktır. Bu çerçevede Türkiye'de 1995-2023 yılları arasında yüzdelik değişimleri ile GSYİH, Madencilik, İmalat, Enerji ve Kimya Sanayi Üretimleri etrafında bir inceleme yapılmıştır. Veri setimizin durağanlığı kontrol edilmiş, gecikme uzunluğu belirlenerek bir VAR tahmini yapılmıştır. Johansen Eşbütünleşme testi ile uzun dönem ilişkileri tespit edilmiş, otokorelasyon ve değişen varyans sorunu olmadığı anlaşıldıktan sonra, Varyans ayrıştırması gerçekleştirilmiştir. Bu sayede belirlenen sektörlerin GSYİH üzerindeki etkileri üzerine çıkarımlarda bulunulmuştur. Analiz sonuçları çerçevesinde GSYİH ve İmalat Sektörü arasında Madencilik, Enerji ve Kimya Sektörlerine nazaran daha güçlü bir ilişki olduğu gözlemlenirken, sektörel bazdaki şokların zamanla azaldığı sonucuna ulaşılmıştır. Ayrıca değişkenler arasında VAR Etki-Tepki Analizi gerçekleştirilmiştir. Bu çerçevede veri setindeki değişkenlerin kısa ve uzun dönemde birbirleri üzerinde etkili oldukları gözlemlenmiştir. İlk dönemlerde şokların daha belirgin olduğu, süreç içerisinde ise ekonominin dengeye ulaştığı sonucu elde edilmiştir. Bu kapsamda ağır sanayi hamlesinin sektörel olarak farklı tepkiler verdiği görülürken, aynı zamanda ekonomik büyüme ve kalkınma açısından öneme sahip olduğu doğrulanmıştır.

## THE IMPACT OF INDUSTRIAL PRODUCTION ON ECONOMIC GROWTH: NEW EMPIRICAL EVIDENCE FOR TÜRKİYE IN A MATERIAL DEVELOPMENT FRAMEWORK

**Abstract**

The economic literature, which has gone through many different processes after the development of humanity and the formation of states, has considered economic growth and development as the most important basis in terms of the goals it has adopted at the basic point. There are different views in the literature regarding economic growth and development. Historically and theoretically, they have aimed to manage this process with different methods and tools. Considering the industrialisation processes of developed countries, the importance of development and growth is a reference for developing countries. Economic growth and development in Turkey has been one of the most emphasised issues since the first years of the Republic. Growth and development hypotheses have been addressed from many different perspectives. It is aimed to examine the view of production with the Heavy Industry Initiative proposed by Prof. Dr. Necmettin Erbakan a crucial part of Material Development in Türkiye. Development plans for five-year periods, which constitute the most important among the different development strategies realised in Turkey's economic history, have been replaced by short-medium and long-term development plans over time. One of the cornerstones of development is the process of economic growth, which Erbakan also calls economic growth. According to Erbakan, this economic growth will enable the country to reach the position of a developed country by ensuring development. In this context, an examination has been conducted on GDP, Mining, Manufacturing, Energy, and Chemical Industry Productions in Türkiye between 1995-2023, focusing on percentage changes. The stationarity of our data set has been checked, and a VAR estimation has been performed by determining the lag length. Long-term relationships have been identified through the Johansen Cointegration test, and after confirming that there are no issues of autocorrelation and changing variance, variance decomposition has been carried out. Thus, inferences have been made regarding the effects of the identified sectors on GDP. Based on the analysis results, it has been observed that there is a stronger relationship between GDP and the Manufacturing Sector compared to the Mining, Energy, and Chemical Sectors, and it has been concluded that sectoral shocks diminish over time. Additionally, VAR Impulse-Response Analysis has been conducted among the variables. In this context, it has been observed that the variables in the dataset have an impact on each other in both the short and long term. It has been concluded that the shocks were more pronounced in the initial periods, while the economy reached a balance over time. In this regard, it has been confirmed that the heavy industry initiative exhibited different responses sectorally, while also being significant in terms of economic growth and development.

**Giriş**

İktisat literatüründe sıkça yer alan büyüme ve kalkınma kavramları çeşitli açılardan incelenerek yeni teori ve modeller üzerinde durularak her zaman güncelliğini korumuştur. Ekonomide gerçekleşecek olan büyümenin kalkınmayı da beraberinde getireceği söylemi birçok iktisatçı tarafından irdelenerek incelenmiştir. Ekonomik büyümenin kişilerin refah düzeylerinde meydana getireceği iyileşme, sermaye birikim, yatırım, üretim, hammadde, teknoloji gibi birçok olguya katkıda bulunarak sağlık, eğitim, lojistik, ulaştırma gibi birçok kalkınma verilerine de destek olacağı görüş birliği oluşmasa da net bir şekilde ifade edilebilecek durumdur.

Ekonomik büyümenin tek başına kalkınmayı getirmeyeceği aynı zamanda aşikar olmakla birlikte aynı zamanda tek başına kalkınmanın da bir sonuca ulaşmakta zorlanacağı düşünülebilmektedir. Gelişmiş ülkelerde tamamlanmış olan kalkınma sürecinin halen devam eden ekonomik büyüme ile de ilişkisi mevcuttur. Bu ülkelerde tamamlanan sanayileşme ile kalkınma süreci de dönemsel olarak tamamlanmıştır. Örneğin gelişmiş ülkelerde tamamlanan kalkınma süreçleri eğitim sayesinde nitelikli iş gücünü de beraberinde getirerek sanayileşmenin vermiş olduğu katkı dolayısıyla da ekonomik büyümeye yardımcı olmaya devam etmektedir. Öte yandan gelişmekte olan ülkeler açısından bir değerlendirme yapıldığın da ise ekonomik büyümenin eşlik ettiği kalkınma süreci devam etmektedir. Kalkınma sürecine katkı veren ekonomik büyüme ise bu ülkelerde devam eden sanayileşme birlikte sürmektedir. Yukarıda verdiğimiz örnek üzerinden ilerleyecek olursak gelişmekte olan ülkelerde de bu durum incelenecek olursa kalkınma sürecinde etkin olamaması nedeniyle nitelikli iş gücü ya da etkin teknolojik ürüne sahip olamamalarından kaynaklı büyüme oranları da etkilenebilmektedir. Bu nokta kalkınma ve büyümenin birbirleri ile etkileşimde olduğu fikri uygun olabilmektedir. Kalkınmanın oluşturulabilmesi amacıyla gerçekleşecek ekonomik büyümenin lokomotif sektörler çerçevesinde ilerlemesi her iki teori açısından da önem arz ettiği ve en çok etkili olabilecek alanlarda aktif olarak gerçekleşebilmesi önemlidir.

Maddi Kalkınma tezi çerçevesinde Erbakan'ının sunmuş olduğu ağır sanayi fikri tam da bu konuda ön plana çıkmaktadır. Makine yapan makine çıkışı ile Erbakan aslında üretim zincirinin temel halkalarını birleştirebilmeyi amaçlayarak kendinin de ifade ettiği şekilde iktisadi büyüme ve gelişimin kalkınmayı sağlayacağına ifade etmiştir. Bu söylev zaman içerisinde çeşitli uygulamalar ve hamlelerle hayata geçirilmeye çalışılmış, bazı dönemlerde güncellenmiş olsa da temel dayanak noktasını yitirmemiştir.

Çalışmamızda Erbakan'ın önermiş olduğu Ağır Sanayi Hamlesi fikri çerçevesinde 1995-2023 yıllarını kapsayan imalat-madencilik-kimya-enerji üretimleri açısından GSYİH ile tespit edilmiştir. Bu sayede maddi kalkınma tezinin ağır sanayi kollarına ait verilerle güncel metotlar çerçevesinde incelemesi gerçekleştirilerek günümüze kadar olan zaman içerisindeki gelişimi incelenmiştir. Erbakan'ın gerçekleştirmiş olduğu söylev çerçevesinde incelemeler sonucunda çıkarımlarda bulunularak, gelecekte yapılacak olan çalışmalar içinde öneriler verilmiştir.

**Ekonomik Büyüme**

Büyüme olgusu ekonomik olarak ele alındığında ülkelerin kendi dinamiklerine özel olarak farklılıklar göstermektedir. Her bir ülke için sermaye birikimi, işgücü, teknoloji ve doğal kaynak miktarları çerçevesinde ayrı olarak değerlendirilebilmektedir. Ekonomik büyümeyi genel olarak üretilen toplam mal ve hizmetlerin artması ya da kişi başına düşen gelir miktarının artırılması olarak açıklanabilmektedir[1]. Bu artışlar ise genelde ülkenin reel GSYİH miktarının ülke nüfusuna bölünmesi ile kişi başına düşen reel GSYİH oranı bulunarak hesaplanabilmektedir. Bireylerin refahını ele alan bu yöntem ekonomik anlamda bir ülkenin performansını değerlendirmekte sıkça kullanılmaktadır. Bu ise ekonomide bulunan tüm paydaşların adil bir şekilde karşılaştırılabilmesini kolaylaştırmaktadır. Öte yandan meydana gelen artışın ise büyüme olarak nitelendirilerek ele alınabilmesi için öncelikle geçici değil sürekli olması gerekmektedir. Bu ise ülke içinde bulunan tüm kaynaklardan verimli ve etkin bir şekilde kullanılması sonucu oluşabilecektir.

Ekonomik büyüme iktisat teorisi açısından farklı iktisat ekolleri ve farklı iktisatçıların temel çalışma alanı olarak karşımıza çıkmaktadır. Bu çalışmaların birçoğunda iktisat literatürüne önemli katkılar yapan birçok ekonomist, ekonomik büyümenin en önemli değişkeni olarak üretimi değerlendirmektedir[2]. Bu durum her ülke için farklılık gösterse de bu çalışmalardan bazıları şu şekildedir: Klasik iktisadi görüş açısından bir değerlendirme yapıldığında A.Smith, T.R.Malthus, F.P. Ramsey, D.Ricardo ve J.A. Schumpeter gibi büyük ekoller akla gelebilmektedir. İngiltere'nin diğer ülkelere kıyasla yaşam standartlarında daha yüksek bir görünüme sahip olmasını irdeleyen Adam Smith, bunı daha çok uzmanlaşma, iş bölümü ve makineleşme ile ilişkili olduğunu aktarmıştır[3]. Öte yandan teknolojide gerçekleşen gelişmeleri

[1, 2] Meliha Nur Köstepen, "İslam İktisadi Perspektifinden Sosyal Adalet ve Kalkınma", *Maruf İktisat İslâm İktisadı Araştırmaları Dergisi*, sy 8 (31 Aralık 2024): 48-60, https://doi.org/10.58686/marufiktisat.1570422.

[3] Şerif Canbay, "Türkiye'de Elektrik Üretimi İçin Kullanılan Petrol Tüketimi, Yenilenebilir Enerji Kullanımı İle İktisadi Büyüme Arasındaki İlişkilerin Analizi", *Ekev Akademi Dergisi, sy 81 (27Şubat 2020):467-488*

ilk olarak büyüme kavramı içerisinde ele alan ise Schumpeter olmuştur[4] (Schumpeter,1939). Marx'a ise sanayileşmenin en büyük faktörü olarak yeni makine ve teçhizatların buluşu doğrultusunda artan makineleşmeyi temel almıştır. Neoklasik iktisat ekolünden olan R.M. Solow ise teknolojide gerçekleşen gelişmelerin, üretim kanalı ile büyümeyi etkilediğini vurgularken Kaldor ise sektörel olarak sanayinin ekonomik büyümenin temel hareket noktası olduğunu vurgulayan Post Keynes ekolünü benimseyen bir iktisatçı olarak karşımıza çıkmaktadır[5]. Temel olarak farklı görüşlere sahip iktisat ekollerinden verilen örneklerden de anlaşılacağı üzere sanayi üretiminin büyüme üzerinde ciddi bir paya sahip olduğu gözlemlenebilmektedir. Tüm bunların yanında; fiziksel ve beşerî sermaye, nüfus, teknoloji, iklim, kültür, kamu politikaları, gelir dağılımında eşitliğin sağlanması gibi daha bir çok faktörün ekonomik büyümeyi etkilediğini vurgulamışlardır[6].

Ekonomide hedeflenen büyüme ülke içerisinde yaşayan insanların refahını iyileştirerek daha iyi yaşam standartları sunabilmek adına önemli bir adım olarak karşımıza çıkmaktadır. Bir ülkede gerçekleşen istihdam, üretim, tüketim vb. olgularda meydana gelen değişiklikler büyüme sürecinin en önemli faktörleri olarak gözlemlenebilmektedir. Ülke içerisinde meydana istihdam artışı işsizlik oranlarının azalmasına, sermaye birikiminde meydana gelen artış üretim miktarının artmasına dolayısıyla da gelirde meydana gelen iyileşmeye öncül olarak makroekonomik açıdan toplumsal refah artışını sağlamaktadır.

**Maddi Kalkınma ve Ağır Sanayi Hamlesi**

Ekonomik kalkınma içinde birçok farklı teori ve görüş literatürde mevcut olmakla birlikte Türkiye özelin Prof. Dr. Necmettin Erbakan'ın sunmuş olduğu Maddi Kalkınma Modeli önem arz etmektedir. Dönemin Türkiye'sinde ithal ikameci politikaların ön planda olduğu Maddi Kalkınma Modeli de bu politikalara uygun olarak gözlemlenebilen Ağır Sanayiye dayalı bir model olarak karşımıza çıkmaktadır. Bu modelin Millî Görüş geleneğiyle ortaya çıkmasının en önemli faktörler arasında siyasi-ekonomik olgular çerçevesinde şekillenmiştir. O dönemlerde Türkiye'de planlı kalkınma modeli uygulanırken birincil amaç olarak sanayileşme görülmekteydi. İkinci dünya savaşı dönemi sonrasında teknoloji transferi sayesinde sanayileşmenin büyümeyi tetikleyeceği düşünülmekteydi. Temel teknoloji transferleri çerçevesi ağır sanayi ürün ve teknolojilerini kapsamaktaydı. Genel olarak teknoloji transferi ile üretken sermayenin üretim araçlarınca kullanılarak ikinci dünya savaşı sonrası dönemde sermaye birikimlerinin iç dönük bir genişleme sergilemesini sağlamaktaydı. Boratav'a göre o dönem Türkiye'nin benimsemiş olduğu strateji de içe dönük dışa bağımlı iktisadi büyüme modeli olarak aktarmaktadır[7]. Türkiye'de ithal ikame politikalarına dayalı modeli daha çok dayanıklı tüketim mallarını içeren montaj sanayi olarak gözlemlenebilmekteydi. Bu noktada Erbakan temelde dışa dayalı olan ekonomik büyümenin yerli üretimle gerçekleştirilmesini düşünmekteydi.

[4] J.A. Schumpeter, *Business Cycles: A Theoretical, Historical, and Statistical Analysis Of The Capitalist Process,* New York and London: Mcgraw-Hill.
[5] Ali Doğdu, "G7 ve E7 Ülkelerinde GSYİH ve Yenilenebilir Enerji Üretimi İlişkisi: Toda-Yamamoto Panel Nedensellik Analizi", *Economics And Management* 3, Sy 2 (2022).
[6] Köstepen, "İslam İktisadi Perspektifinden Sosyal Adalet ve Kalkınma".

Maddi kalkınma çerçevesinde temel argüman ve birincil şart olarak ağır sanayi üretimlerinin gerçekleştirilebilmesidir. Erbakan ağır sanayiyi fabrikayı kuran fabrika-makine yapan makine olarak nitelendirmiş ve temel yatırım mallarının üretildiği sanayi olarak açıklamıştır[8]. Ağır sanayi hamleleri bölgelere göre uygulanabilirse ülkenin topluca kalkınması sağlanabilecektir. Erbakan ağır sanayi içerisinde üretim amacıyla kullanılan teknoloji ve araçların ithalata dayalı olmaktan ziyade yerli üretim ile gerçekleştirilmesini hatta bu ürünlerin üretimin sağlayacak teknolojilerinde yerli olması görüşünü savunmaktaydı. Ağır sanayi 1970'li yıllarda temel üretim teknolojisi ürünleri olarak benimsenmişken ağır sanayide kullanılacak teknolojik araçlarında üretimi için teknolojik alt yapı ve üretimi sahip olunması gerektiğini vurgulamaktaydı. Ağır sanayi kavramı Erbakan'ın siyasi söylev ve programlarında sıkça yer almış, siyasi yaşamının başladığı dönemden itibaren Türkiye'nin azgelişmiş, mali yapı çerçevesinde düşük, yüksek enflasyon oranına sahip, yabancı sermaye etkisi altında ve dışa bağımlı kalan bir sanayi üretimi içerisinde bulunduğunu aktarmıştır. Kendi önermiş olduğu kalkınma söyleminde sıkça Liberal-Kapitalist-Sosyalist sistemleri eleştirmiştir. Ağır sanayi hamlesi dayanağı ve kaynağı, sanayileşmeye verilen önem ve uygulamadaki hatalar son olaraksa uygulanabilirliği ile sanayileşmenin tamamlanması olmak üzere üç aşamaya ayırmıştır[9].

Erbakan Maddi Kalkınma Modeli çerçevesinde Ağır Sanayi oluşumunu sanayi, enerji, madencilik, sulama ve ulaştırma olarak beş farklı şekilde kategorize etmiştir. Planlamasını ise; 218 adet büyük sanayi tesisi, 29 adet büyük enerji tesisi, 29 adet madencilik tesisi, 58 adet büyük sulama tesisi 49 adet büyük ulaştırma tesisi olmak üzere toplamda 383 proje olarak gerçekleştirmiştir. Büyük sanayi tesislerinin içeriğinde; zorunlu ihtiyaçlara yönelik tesisler, ağır sanayi tesisleri, ağır harp sanayi tesisleri, elektrik ve telekomünikasyon ve yaygın sanayi kuruluşlarını içermektedir. Büyük Enerji tesislerinin içeriği ise; Nükleer/Atom, Hidroelektrik, Termik ve Rafine tesislerini içermektedir. Büyük Madencilik Tesislerini ise; Cevher İstihracı, Zenginleştirme ve İzabe olarak belirlenmiştir. Büyük Sulama Tesisleri ise; Barajlar, Göletler, Derin Kuyular ve Sulama alanları olarak açıklanmıştır. Büyük ulaştırma tesislerinin içeriği ise; Otoyollar, Bölünmüş Yollar, Hızlı Tren, Hava Limanları ve Limanlar olarak açıklanmıştır[10]. Türkiye'nin çokça zengin doğal kaynaklara sahip bir ülke olduğunu ve bunun ağır sanayi için elzem olduğunu aktararak ağır sanayi hamlesinin Türkiye için faydalı sonuçlar veren bir olgu olarak gözlemlenebileceğini iletmiştir. Bu noktada devlete ise düzenleyici ve etkin bir rol alarak ağır sanayi için lokomotifi oluşturması gerektiğini ifade etmiştir. 1980 sonrasına kadar devletin piyasayı destekleyici bir rol alması gerektiğini bunun yanında özel sektöründe yatırımlarını genişleterek bu sürece katkı sağlaması gerektiğini aktarmıştır. Keynesyen birikim modelinin hâkim görüş olduğu bu dönemlerde bu söylemi yakın görüşlerde bulunmuşlardır. Tüm bu düşüncelerini Erbakan siyasi hayatı boyunca kurmuş ve içinde bulunmuş olduğu politik oluşumlarda çeşitli söylev ve sloganlarla dile getirmiştir. 1987 yılında Brundtland raporunda ele alınan sürdürülebilir kalkınma söylevi Erbakan için 1997 yılı sonrasında ağır sanayi hamlesi, sanayide milli hamle gibi benimsenmeye başlanmıştır. 1998'de Erbakan parti çalışma programına sürdürülebilir kalkınmayı eklemiştir. Burada sürdürülebilir kalkınma için doğayı koruyan, insan sağlığına önem verecek şekilde doğal kaynakların kullanılarak ekonomik kalkınma ve büyümenin gerçekleştirilmesi gelecek nesillerin doğal ve sosyal haklarına sahip çıkarak çevre politikalarının gözetileceği aktarılmıştır. Bu süreç sırasında Erbakan'a çeşitli eleştirilerde bulunulmuş, özellikle seçim beyannamelerinde serbest piyasa, özel mülkiyet, dışa açıklık, rekabet ve üretim gücünün yüksekliği çerçevesinde oluşturdukları sürdürülebilir kalkınma politikaları nedeniyle, liberal söylemlere yönelinmesi temel eleştirilerin odak noktasını oluşturmaktadır.

---

7 Mehdi Pekedis, "Millî Görüş Hareketinde Kalkınma Söylemi: Tarihsel ve İdeolojik Bağlamda Bir Okuma", *Ekonomi Yönetim Politika 2,* 1.(25 Haziran 2024).

8 Nazli Akpınar. *Maddi-manevi kalkınma: Necmettin Erbakan örneği,* Yüksek Lisans Tezi, Marmara Üniversitesi, 2009.

9 İrfan, *Ersin. Necmettin Erbakan'ın ekonomik söylem ve uygulamaları,* Yüksek Lisans Tezi, Abant İzzet Baysal Üniversitesi, 2015.

Öte yandan Erbakan'ın Maddi Kalkınma Modeli ve Ağır Sanayi Hamlesi iktisat literatüründe yer alan bazı modellerle bazı noktalarda benzeşmektedir. Yanlış Paradigma Modeli çerçevesinde Türkiye'ye gelen bazı yabancı uzmanların Tarımsal Üretime öncelik verilmesi üzerine kurulmuş raporlar nedeniyle Türkiye'nin Cumhuriyet Dönemi kalkınmasında engel oluşturduğu şeklinde eleştirilerde bulunan Erbakan bu eleştirileri çerçevesinde Bağımlılık Teorisiyle uyum sağlamıştır. Friedrich List'in Bebek Endüstri tezi ile Erbakan'ın sanayileşme düşüncesi de benzerlik göstermektedir. Erbakan'a göre sanayileşmenin ilk evrelerinde devlet korumacı bir rol üstlenmeli (ithalat kısıtlamaları-İç Piyasaya dönük üretim) daha sonra sanayileşme süreci tamamlanınca serbest piyasayı desteklemelidir. Gerschenkron geç kalkınma teorisindeki gibi Erbakan da sanayileşme süreciyle birlikte devlet destekli bir şekilde kalkınmanın olabileceği fikrindedir. Öte yandan kalkınma için Erbakan'ın önerdiği bölgesel kalkınma şirketleri düşüncesi de Endojen kalkınma modeli ile özdeşleşebilmektedir.

[10] İrfan Ersin, *Necmettin Erbakan'ın Ekonomik Söylem ve Uygulamaları*.

**Veri, Metodoloji ve Ampirik Bulgular**

Bu çalışmada Prof. Dr. Necmettin Erbakan'ın Maddi Kalkınma Modeli çerçevesinde Ağır Sanayi Hamlesi söylevi incelenecektir. Bu amaçla 1995-2023 yılları arasında Türkiye için yüzdelik değişimleri çerçevesinde GSYİH, Madencilik, İmalat, Enerji ve Kimya Sanayi Üretimleri çerçevesinde bir incelme gerçekleştirilmiştir. Tablo 1'de verildiği üzere veri setine ait değişkenler TÜİK ve Dünya Bankasından elde edilmiştir.

**Tablo 1.** Değişkenlerin Tanımlanması

| **Değişkenler** | **Açıklama** | **Kaynak** |
|---|---|---|
| İmalat | İmalat Sanayi % Üretim Oranları | TÜİK |
| Maden | Maden Sanayi % Üretim Oranları | TÜİK |
| Kimya | Kimya Sanayi % Üretim Oranları | TÜİK |
| Enerji | Enerji % Üretim Oranları | Dünya Bankası |
| GSYİH | GSYİH % Oranları | Dünya Bankası |

Çalışmamızda analiz yöntemi olarak Vektör Otoregresif Modeli (VAR) kullanılmıştır. Bu model, ekonometrik zaman serisi analizlerinde sıkça kullanılarak makroekonomik bir değişkenin diğer değişkenler üzerindeki dinamik ilişkilerini inceleyen istatistiki bir model olarak literatüre dahil olmuştur[11]. VAR modeli analizlerinde kullanılacak değişkenlerin dışsallık açısından bir sonucun netleştirilemediği durumda kullanılır ve 1980'de Sims'in geliştirmiş olduğu bir modeldir[12]. Bu model yapısal kısıtlama olmadan, seçili değişkenleri bütün halinde değerlendirerek bir sistem bütünlüğüne sahip olarak analiz edebilmektedir[13].

$$y_t = \alpha + \sum_{i=1}^{t} \beta_i y_{t-i} + \sum_{i=1}^{t} \gamma_i x_{i-1} + \varepsilon_{1t} \qquad (1)$$

[11] Lovrinovic, Igor, and Maja Benazic. 'A VAR Analysis of Monetary Transmission Mechanism in the European Union.' *Zagreb International Review of Economics and Business 7*, no. 2 (2004): 27-42.", https://hrcak.srce.hr/file/56300.
[12] Ali Doğdu, "Türkiye'de Enflasyon Hedeflemesi ve Genişletilmiş Taylor Kuralının 2008-2023 Yılları Arasında Geçerliliğinin Test Edilmesi", *Toplum Ekonomi ve Yönetim Dergisi* 4, sy Özel (29 Ekim 2023): 166-83, https://doi.org/10.58702/teyd.1357546.
[13] Ferhat Başkan Özgen ve Bülent Güloglu, "Türkiye'de İç Borçların İktisadî Etkilerinin VAR Tekniğiyle Analizi", Haziran 2004, https://open.metu.edu.tr/handle/11511/58514.

$$y_t = \alpha + \sum_{i=1}^{t} \beta_i X_{t-i} + \sum_{i=1}^{t} \gamma_i y_{i-1} + \varepsilon_{1t} \qquad (2)$$

Standart VAR modeli içerisinde bulunan tüm değişkenlerin hem kendilerinin hem de diğer değişkenlerin gecikmeli değerleri üzerinden tanımlama yaptığı çok boyutlu basit bir zaman serisinin perspektifi olarak ele alınmaktadır[14]. Yukarıda verilen denklem 1 ve 2 de yer alan Ɛ (rassal hata) terimini temsil eder ve bu sayede sistemde oluşan şokları göstermektedir[15]. Örneğin; *yt* ve *zt* benzeri iki farklı zaman serisinin olduğu bir modelde *yt*'nin zaman içinde geçekleştirdiği hareket, *zt*'yi şu anki ve geçmişteki değerlerini yine benzer şekilde *zt*'nin zaman içinde gerçekleşen bir hareketinin de *yt*'nin şimdi ve geçmişteki varyasyonlarını etkileyebildiği gözlemlenmektedir. Bu çerçevede iki değişkenli basit bir sistemi aşağıdaki gibi ifade edebiliriz:

$$y_t = b_{10} - b_{12} z_t + y_{11} y_{t-1} + \varepsilon_{yt} \qquad (3)$$

$$z_t = b_{20} - b_{21} y_t + y_{21} y_{t-1} + y_{22} z_{t-1} + \varepsilon_{yt} \qquad (4)$$

3 ve 4 numaralı denklemlerde; *yt* ve *zt* değerlerinin durağanlıklarının var olduğu *εyt ve εzt*'nin $\sigma y$ ve $\sigma z$ sahip oldukları standart sapmaları ile beyaz gürültü (white noise) oldukları $\{\varepsilon yt\}$ ve $\{\varepsilon zt\}$'nin ilişkilerinin olmadan white noise hata terimlerine sahip oldukları varsayımsal olarak ele alınmaktadır. Makroekonomik modellerde bulunan değişkenlerin genellik kendi aralarında dinamik geriye dönük bir beslemeye sahip oldukları aktarılmaktadır. Sistemde herhangi bir zaman serisinin, zaman içerisinde izlediği trendin sistemde bulunan diğer serinin trendinden bağımsız hareket edip etmediğini bilmemekteyiz. Buna benzer simetrik ilişkilerin olduğu zaman serilerinde çok denklemli sistemler için VAR yöntemi ile inceleme yapılabilmektedir[16]. Bu modeller birden fazla denklemlerle oluşan zaman serisi modellemelerinde dışsal ve içsel ayrım yapılma gerekliliği ortadan kalkarak tüm değişkenler içsel olarak kabul edilebilmektedir.

Zaman serisi çerçevesinde gerçekleştirilen ekonometrik analizlerde birim kök testinin analiz edildiği en temel yöntemler 1979 Dickey ve Fuller'in geliştirmiş olduğu ADF birim kök testi ile 1988 Philips ve Perron PP birim kök testi olarak karşımıza çıkmaktadır[17]. Değişkenlerin durağanlıkları veri setinin zaman içerisinde her bir gecikme dönemi için belirli bir değere doğru yakınlaşarak sabit ortalama, sabit varyans ve gecikme seviyesi ile bağlı olarak kovaryansının bulunması olarak ifade edilebilmektedir[18]. Çalışmamızda durağanlık birim kök testi sonuçları aşağıda Tablo 2'de verilmiştir.

---

[14] Doğdu Ali, "Türkiye'de Enflasyon Hedeflemesi ve Genişletilmiş Taylor Kuralının 2008-2023 Yılları Arasında Geçerliliğinin Test Edilmesi".
[15] Christopher A. Sims, *Macroeconomics and Reality*, Econometrica 48, sy 1 (1980): 1-48, https://doi.org/10.2307/1912017.
[16] Kazdağlı, Hüseyin. "Türkiye Cumhuriyeti Merkez Bankası'nın Kuruluş Tarihçesi ve 1934-1938 Dönemindeki Para Politikasının VAR Yöntemiyle Analizi." *Hacettepe Üniversitesi İktisadi ve İdari Bilimler Fakültesi Dergisi* 14, no. 2 (1996): 35-52.
[17] Doğdu, Ali. *Taylor kuralının gelişmekte olan ülkeler üzerindeki geçerliliğinin Dumitrescu Hurlin panel nedensellik analizi ile test edilmesi*, Yüksek Lisans Tezi, Konya Gıda ve Tarım Üniversitesi, (2019).

**Tablo 2.** Panel Birim Kök Testi Sonuçları

| Model | ADF | | PP | |
|---|---|---|---|---|
| | **İstatistik Değeri** | **Olasılık Değeri** | **İstatistik Değeri** | **Olasılık Değeri** |
| Sabitli | 95.8903 | 0.0000 | 105.473 | 0.0000 |
| Trendli ve Sabitli | 87.0868 | 0.0000 | 134.191 | 0.0000 |
| Trendsiz ve Sabitsiz | 69.1690 | 0.0000 | 979.629 | 0.0000 |

Yukarıda Tablo 2'de verilen ADF ve PP birim kök testi analiz sonuçları çerçevesinde verilerimizin seviyede durağan oldukları görülebilmektedir. Burada ADF ve PP birim kök testi sonuçlarının da tutarlı olduğu anlaşılabilmektedir.

Öte yandan tahmin edilecek VAR Modeli için optimum gecikme uzunluğunun belirlenebilmesi için yapılan analiz sonuçları Tablo 3'te verilmiştir.

**Tablo 3.** VAR Gecikme Uzunluğunun Belirlenmesi

| Gecikme | LogL | LR | FPE | AIC | SC | HQ |
|---|---|---|---|---|---|---|
| 0 | -287.2787 | NA | 3998.929 | 22.48298 | 22.72492 | 22.55265 |
| 1 | -258.1430 | 44.82417 | 3034.453 | 22.16485 | 23.61650 | 22.58287 |
| 2 | -232.3056 | 29.81236 | 3621.400 | 22.10043 | 24.76179 | 22.86681 |

VAR modelinin tahmin edilebilmesi için Tablo 3'te verilen sonuçlara göre en uygun gecikme sayısı belirlenmiştir. Bu çerçevede Akaike Bilgi Kriteri çerçevesinde en düşük değere sahip olan ikinci gecikme uzunluğu tercih edilmiştir. Modelimiz bundan sonraki süreçte VAR tahmini gerçekleştirilirken 2 gecikme uzunluğu dikkate alınacaktır.

Aşağıda Şekil 1'de ise oluşturulan VAR Modeline ait istikrar koşulunu gösteren Ters Köklerin Modülüs Grafiği verilmiştir.

[18] "Doğdu, Ali. *Taylor kuralının gelişmekte olan ülkeler üzerindeki geçerliliğinin Dumitrescu Hurlin panel nedensellik analizi ile test edilmesi.*

**Şekil 1.** VAR Modeli Ters Köklerin Modülüs (İstikrar Koşulu) Grafiği

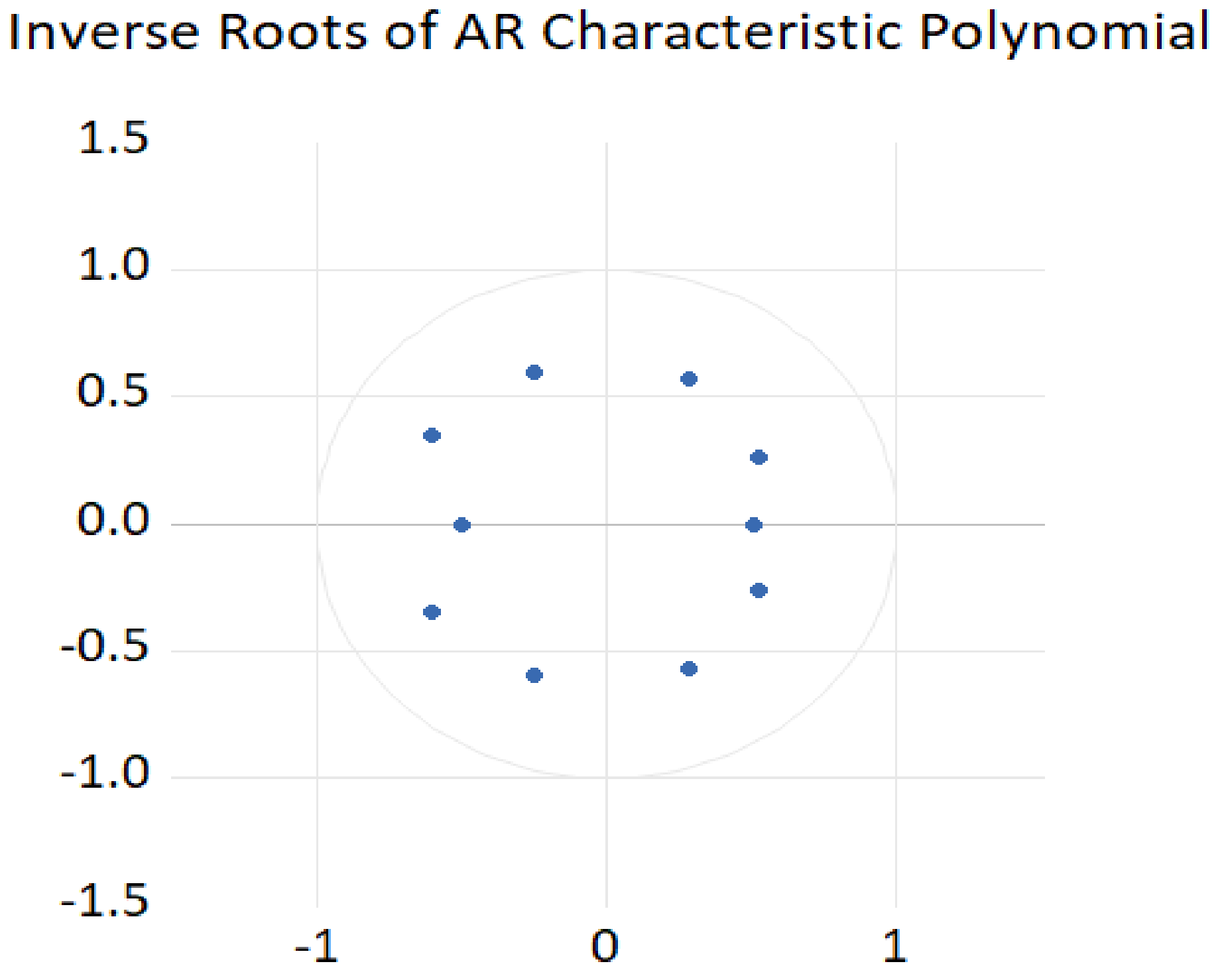

Şekil 1'de verilen grafikten anlaşılacağı üzere ters köklerin herhangi bir sorun olmaksızın tamamının birim çember içinde olduğu ve 1'den küçük bir değere sahip olduğu anlaşılmaktadır.

**Tablo 4.** VAR Kalıntı Seri Korelasyon LM Testi Sonuçları

| Lag | LRE* stat | df | Prob. | Rao F-stat | df | Prob. |
|---|---|---|---|---|---|---|
| 1 | 31.63721 | 25 | 0.1688 | 1.349769 | (25, 46.1) | 0.1858 |
| 2 | 27.02157 | 25 | 0.3548 | 1.103809 | (25, 46.1) | 0.3763 |
| *H0:h gecikmede seri korelasyon yoktur* | | | | | | |
| Lag | LRE* stat | df | Prob. | Rao F-stat | df | Prob. |
| 1 | 31.63721 | 25 | 0.1688 | 1.349769 | (25, 46.1) | 0.1858 |
| 2 | 55.77152 | 50 | 0.2668 | 1.102219 | (50, 35.3) | 0.3850 |
| *H0:1'den h'ye kadar olan gecikmelerde seri korelasyon yoktur* | | | | | | |

Tablo 4'te verilen sonuçlarda da görüldüğü üzere modelimizin kalıntılarında herhangi bir Seri Korelasyon sorunu olmadığı anlaşılmıştır.

**Tablo 5.** VAR Modeli Artıklarında Heteroskedastisite Sonuçları

| Chi-sq | df | Prob. |
|---|---|---|
| 155.5795 | 150 | 0.3607 |

Tablo 5'te verilen sonuçlarda ise oluşturulan VAR modelimizde kalıntılarda değişen varyans sorunu olmadığı gözlemlenmiştir.

Birden fazla seriden meydana gelen denklemler için eşbütünleşmenin varlığı uzun dönemde ve vektörel olarak analiz edilir. Tüm değişkenler arasında bir eşbütünleşme ilişkisi varsa, veriler arasında uzun dönemli bir ilişkinin olduğu anlaşılmaktadır[19]. Johansen eşbütünleşme testinin temelinin vektörel bir oluşum olması ile VAR (Vektör Otoregresif) bir modelde eşbütünleşmede bulunan vektörlere ait değerlerin anlamlığı İz ve Maksimum Özdeğer ile analiz edilebilmektedir. Johansen ve Juselius (1990) tarafından bu testler için oluşturulan denklemler ise 5 ve 6 olarak verilmiştir[20];

$$iz\ istatistiği = -T \sum_{i=r+1}^{p} \ln(1 - \mu_{r+1}) \qquad (5)$$

$$maksimum\ \text{öz}\ değer\ istatistiği = -T \ln(1 - \mu_{r+1}) \qquad (6)$$

**Tablo 6.** Johansen Eşbütünleşme Sonuçları

| Eşbütünleşme | Özdeğer | İz İstatistiği | Kritik Değer | Olasılık Değeri |
|---|---|---|---|---|
| Yoktur | 0.852411 | 98.04050 | 69.81889 | 0.0001 |
| En az 1 | 0.557516 | 46.02958 | 47.85613 | 0.0735 |
| Eşbütünleşme | Özdeğer | Max. Özdeğer | Kritik Değer | Olasılık Değeri |
| Yoktur | 0.852411 | 49.74645 | 33.87687 | 0.0003 |
| En az 1 | 0.556673 | 21.14966 | 27.58434 | 0.2673 |

Tablo 6'da verilen analiz sonuçları çerçevesinde; İz istatistiği için H0 hipotezi eşbütünleşmenin bulunmadığı (r=0), h1 hipotezi ise en az 1 (r≤1) eşbütünleşme ilişkisine ait varsayımı göstermektedir. Görüldüğü üzere iz istatistiği için Tablo 6'da verilen olasılık değeri karşılığının en az 1 eşbütünleşme ilişkisini verdiği %5 anlamlılık düzeyinin elde edilen kritik değerden büyük olduğunu göstermektedir. Bu durumda en az 1 eşbütünleşme olduğu anlaşılmaktadır. Maksimum özdeğer içinse H0 hipotezi eşbütünleşmenin olmadığı, H1 hipotezi ise en az 1 eşbütünleşmenin olabileceği varsayımını vermektedir. Bu çerçevede de yine aynı şekilde en az

[19] Murad Kayacan ve Ali Doğdu, "Enflasyon Oranı, KDV Gelirleri ve Mevduat Faiz Oranları Bağlamında Para ve Maliye Politikasının Araç Ve Etkilerine İlişkin Yeni Kanıtlar", *Akademik Hassasiyetler* 11, sy 25 (24 Ağustos 2024): 105-35, https://doi.org/10.58884/akademik-hassasiyetler.1445402.
[20] Johansen, Søren, and Katarina Juselius. "Maximum Likelihood Estimation and Inference on Cointegration: With Applications to the Demand for Money." *Oxford Bulletin of Economics and Statistics* 52, no. 2 (1990): 169-210. https://doi.org/10.1111/j.1468-0084.1990.mp52002003.x.

1 eşbütünleşmenin olduğu sonucuna ulaşılabilmektedir. İz ve Maksimum Öz Değer tablosu sonuçlara bize her iki durumda da en az 1 eşbütünleşmenin bulunduğunu göstermektedir.

Şekil 2'de verilen grafikler incelendiğinde GSYİH değişkeninin İmalat sanayi üretiminden kaynaklanan şoklar karşısında vermiş olduğu tepkiler gözlemlenmektedir. Buna göre; İmalat sanayinden gelen bir şoka 1. ve 5. dönemlerde pozitif, 2.-3. ve 4. dönemlerde negatif, diğer dönemlerde ise daha az tepki verdiği ve zaman içerisinde dengelendiği gözlemlenmiştir. İmalat sanayi üretiminde gerçekleşen şoklara, genişleme dönemlerinde pozitif ve daralma dönemlerinde ise negatif GSYİH değişkeninin kayıtsız kalmadığı ve duruma tepki verdiği anlaşılmaktadır. Kimya sanayi değişkeninden kaynaklanan şoklara GSYİH değişkeninin 1. ve 6. Dönem arasında negatif-pozitif tepkiler verdiği daha sonraki dönemlerde ise nispeten bu tepkilerin azalarak 8. Dönemden itibaren dengeye ulaştığı gözlemlenebilmektedir. Maden sanayinde gerçekleşen şoklara ise ilk dönemde negatif, ikinci dönemde pozitif tepki verdiği diğer dönemlerde de tepkinin daha düşük olmakla birlikte devam ettiği 5. Dönemden sonra ise dengelendiği görülebilmektedir. Öte yandan Enerji sanayi üretiminden kaynaklanan şokların GSYİH üzerinde incelenen yıllar bazında genelde dengeli bir oluşumda kaldığı dönem-dönem negatif ve pozitif anlamda küçükte olsa tepkiler verdiği anlaşılabilmektedir.

**Şekil 2.** Genelleştirilmiş Etki-Tepki Fonksiyonları Grafiği

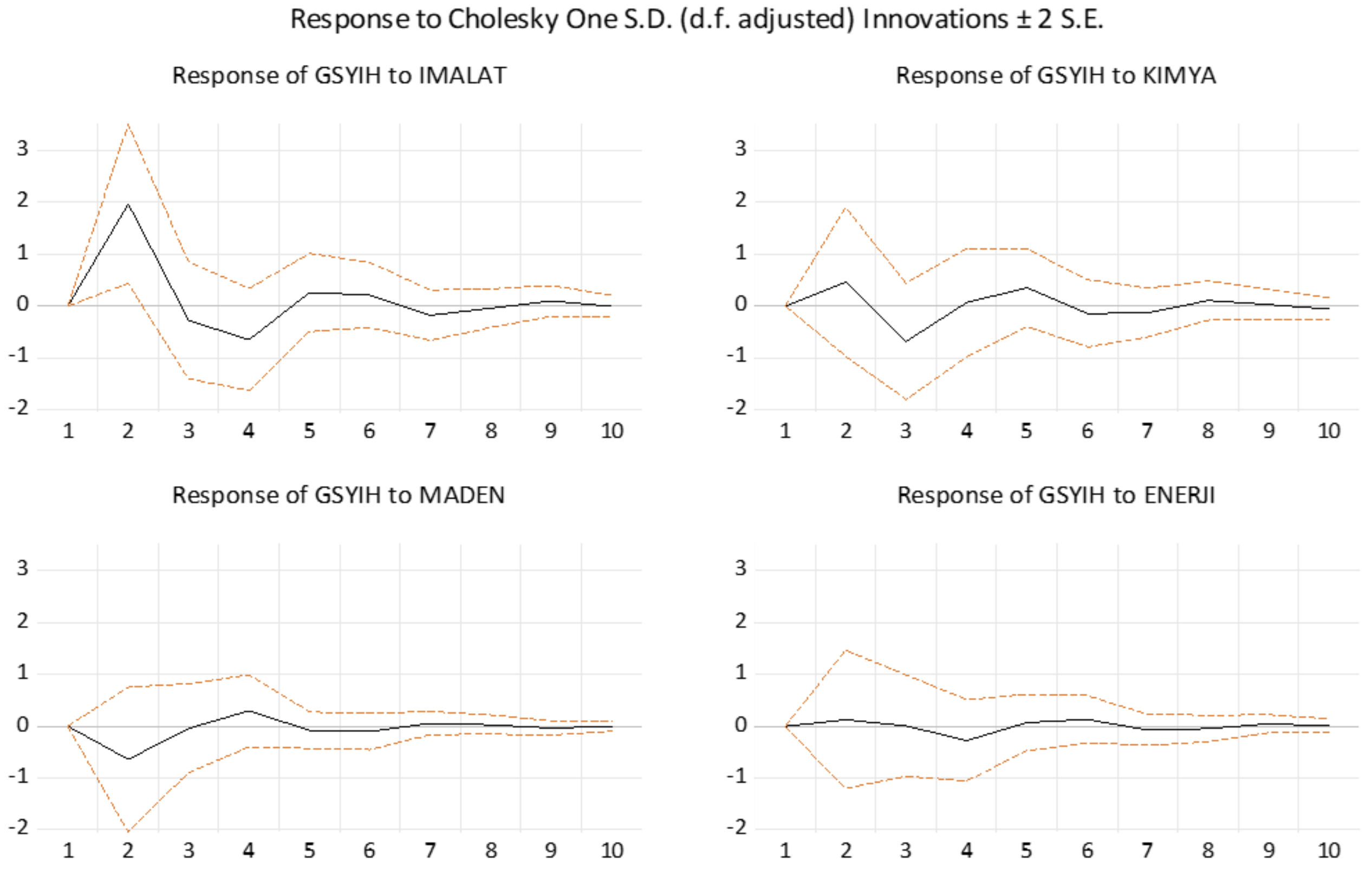

Çalışmamızın önemli bir kısmını oluşturan bir diğer inceleme ise Varyans Ayrıştırması yöntemi kullanılarak gerçekleştirilmiştir. Varyans Ayrıştırması yönteminin tercih edilmesinin en önemli nedeni olarak her bir değişkenin hata terimlerinde meydana gelecek şokların diğer değişkenler tarafından açıklanabilme oranını vermesi ve bu sayede anlamlı bir ilişkinin kurulabilmesi amaçlanmıştır. Tablo 7'de verilen analiz sonuçlarına ele alındığı zaman GSYİH değişkeninin kendisinden kaynaklanan bir şokun ilk dönemde tamamen kendisi ile açıklayabildiği gözlemlenmiştir. İmalat sanayi üretiminde yaşanan değişikliklerin GSYİH hasıla değişkenini 2.dönemden itibaren açıkladığı ve en yüksek değerle ilk dönemde %19,05 olarak belirlendiği görülmektedir. Diğer dönemler incelendiğinde imalat sanayi üretimi değişkeninin en düşük seviyede yaklaşık olarak %17,57 olarak GSYİH değişkenini açıklayabildiği diğer dönemlerde ise genelde %18 seviyelerinde olduğu anlaşılmaktadır. Kimya sanayi üretiminin de 2.dönemden itibaren GSYİH değişkeni üzerinde anlamlı açıklanabilir oranı verdiği görülürken ikinci dönemde en düşük seviye açıklama oranıyla yaklaşık olarak %1,05 diğer dönemlerde ise yaklaşık olarak %3 seviyelerinde gerçekleştiği en yüksek açıklama oranınınsa 10.dönemde %3,59 olarak gerçekleştiği elde edilen sonuçlardan anlaşılmıştır. Maden sanayi üretiminin GSYİH değişkenini 2.dönemden itibaren açıklayabildiği, bu dönemler arasında en yüksek açıklama oranının yaklaşık olarak %2,18 ile 9.dönemde en düşük açıklama oranınınsa yaklaşık olarak %1,82 ile 3.dönemde gerçekleştiği diğer dönemlerde %2,17 seviyelerinde seyrettiği anlaşılmaktadır. Enerji üretiminin GSYİH değişkenini diğer değişkenlere oranla daha az seviyelerde açıkladığı gözlemlense de en düşün seviye açıklanması %0,07 oranı ile 3.dönemde, en yüksek açıklama oranı ise %0,52 ile 10.dönemde gerçekleştiği anlaşılmıştır.

**Tablo 7.** GSYİH Varyans Ayrıştırması Sonuçları

| Dönem | GSYIH | IMALAT | KIMYA | MADEN | ENERJI |
|---|---|---|---|---|---|
| 1 | 100.0000 | 0.000000 | 0.000000 | 0.000000 | 0.000000 |
| 2 | 77.81453 | 19.05080 | 1.049068 | 2.003401 | 0.082206 |
| 3 | 77.53243 | 17.56595 | 3.008680 | 1.818262 | 0.074683 |
| 4 | 75.58854 | 18.90486 | 2.948184 | 2.156217 | 0.402205 |
| 5 | 75.25938 | 18.78074 | 3.409219 | 2.135836 | 0.414827 |
| 6 | 74.96752 | 18.89482 | 3.480514 | 2.168777 | 0.488362 |
| 7 | 74.84026 | 18.94315 | 3.534491 | 2.173862 | 0.508234 |
| 8 | 74.80010 | 18.93132 | 3.577662 | 2.175280 | 0.515641 |
| 9 | 74.76409 | 18.95390 | 3.578409 | 2.179307 | 0.524296 |
| 10 | 74.75788 | 18.94787 | 3.591322 | 2.178606 | 0.524323 |

## Sonuç

Bu çalışmamızda sanayiye dayalı Prof. Dr. Necmettin Erbakan'ın önermiş olduğu Maddi Kalkınma Modeli çerçevesinde Ağır Sanayi Hamlesinin ekonomik büyüme üzerindeki etkileri incelenmiştir. Yapılan analizler çerçevesinde Erbakan'ın söylevlerine bağlı olarak kalkınmanın temel bileşeninin iktisadi büyüme olduğu çıkış noktasından hareketle, ağır sanayi üretimi içerisinde yer alan sanayi sektörlerine ait veriler derlenerek ekonomik büyüme üzerindeki etkileri araştırılarak çıkarımlar da bulunulmuştur.

Erbakan'ın sunmuş olduğu Maddi Kalkınmanın Ağır Sanayi ile oluşturulabilmesi gerçeği irdelenerek yapılan analizler çerçevesinde dönem içerisinde Türkiye için büyüme oranları üzerinde ciddi bir paya sahip olduğu gözlemlenmiştir. Öyle ki tüm değişkenlerin oluşturduğu etki-tepki analizlerinden de anlaşılacağı üzere enerji dışında kalan tüm sektörlerde GSYİH büyümesi üzerinde negatif ve pozitif etkiler gözlemlenebilmiş, enerji sektöründe ise daha düşük oranda bir tepkinin oluşmasının temelde Erbakan'ın da vurguladığı gibi Türkiye'nin dışa bağımlı bir konumda olmasının etkisi olabileceği düşünülebilir.

Yaptığımız diğer analize göre ise GSYİH'nın diğer değişkenler deki şoklara verdiği tepkilerin açıklama oranları incelenmiştir. Bu çerçevede de analiz sonuçları GSYİH'nin diğer değişkenler açısından yaklaşık olarak %25-26'sını açıklayabildiği noktasında sonuçlar vermiştir. Bu nokta da ise Erbakan'ın düşüncesi etrafında ekonomik büyümeyi bu sektörler bazında ele aldığımızda yaklaşık dörtte birini karşılayabildiği yönünde bir çıkarımda bulunabiliriz. Bazı sektörlerde kalan düşük oranların ise aslında analizimize dahil etmediğimiz diğer sektörlerden kaynaklı olabileceği de unutulmamalıdır. Turizm, Tarım, Eğitim, Sağlık ve Askeri alanlarda gerçekleşen çıktı düzeyleri bu çalışmamızda yer almamıştır. Yapılan analizler çerçevesinde gerçekleşen tepki ve açıklama oranlarının Erbakan'ın ağır sanayi hamlesinin etkin bir şekilde işleyebileceğini istatistiki olarak göstermektedir.

Gelecekte yapılacak olan çalışmalar için sektör derinliğinin ve çıktı düzeylerinin detaylandırılarak ele alınması, diğer sektörlerinde analizlere eklenerek net fark düzeyinin incelenebilmesini bu sayede de farklı çıkarımların yapılabilmesini olanaklı kılacaktır.

## Kaynakça